\def\year{2020}\relax
\documentclass[letterpaper]{article} 
\usepackage{aaai20}  
\usepackage{times}  
\usepackage{helvet} 
\usepackage{courier}  
\usepackage[hyphens]{url}  
\usepackage{balance}
\usepackage{graphicx} 
\usepackage{graphicx}  
\usepackage{graphicx}
\usepackage{subcaption}
\usepackage{tikz}
\usepackage{pgfplots}
\usepackage{pgfplotstable}
\usepackage{url}

\nocopyright

\pgfplotsset{
SmallBarPlot/.style={
    font=\small,
    ybar,
    width=\linewidth,
    ymin=0,
    xtick=data,
    xticklabel style={text width=0.8cm, align=center},
    xtick pos=left,
},
BlueBars/.style={
    fill=MidnightBlue!30, bar width=1.5
},
RedBars/.style={
    fill=red!40, bar width=2
}
}

\pgfplotsset{
MediumBarPlot/.style={
    font=\small,
    ybar,
    width=\linewidth,
    ymin=0,
    xtick=data,
    xticklabel style={text width=2.5cm, align=center},
    xtick pos=left,
},
BlueBars/.style={
    fill=MidnightBlue!50, bar width=4.5
},
RedBars/.style={
    fill=red!40, bar width=4.5
}
}
\pgfplotsset{grid style={dashed,gray}}
\pgfplotsset{minor grid style={dashed,red}}
\pgfplotsset{major grid style={dotted,green!50!black}}
\pgfplotsset{compat=1.6}
\usetikzlibrary{scopes}
\usetikzlibrary{calc,chains,positioning,shapes.geometric,shadows}

\pgfplotsset{select coords between index/.style 2 args={
    x filter/.code={
        \ifnum\coordindex<#1\fi
        \ifnum\coordindex>#2\fi
    }
}}

\title{Efficient Data Management for Intelligent Urban Mobility Systems}

\author{
    \textbf{Michael Wilbur \textsuperscript{\rm 1},
    Philip Pugliese \textsuperscript{\rm 2},
    Aron Laszka \textsuperscript{\rm 3},
    Abhishek Dubey \textsuperscript{\rm 1}} \\
    \textsuperscript{\rm 1}Vanderbilt University, Department of Electrical Engineering and Computer Science \\
    \textsuperscript{\rm 2}Chattanooga Area Regional Transportation Authority (CARTA) \\
    \textsuperscript{\rm 3}University of Houston, Department of Electrical Engineering and Computer Science \\
}

\begin{document}

\maketitle

\begin{abstract}

Modern intelligent urban mobility applications are underpinned by large-scale, multivariate, spatiotemporal data streams. Working with this data presents unique challenges of data management, processing and presentation that is often overlooked by researchers. Therefore, in this work we present an integrated data management and processing framework for intelligent urban mobility systems currently in use by our partner transit agencies. We discuss the available data sources and outline our cloud-centric data management and stream processing architecture built upon open-source publish-subscribe and NoSQL data stores. We then describe our data-integrity monitoring methods. We then present a set of visualization dashboards designed for our transit agency partners. Lastly, we discuss how these tools are currently being used for AI-driven urban mobility applications that use these tools.

\end{abstract}

\section{Introduction}

With the proliferation of affordable sensor and networking units, urban mobility systems are entering the connected world at a rapid pace. This has led to a proliferation of research related to improving the experience of using public transportation. As use of public transit continues to increase, transit agencies are looking to AI and machine learning to make existing systems more efficient and thus maximize existing infrastructure. 

AI methods in this context require data from a variety of real-time streams from a variety of sources. For example, traffic prediction and transit optimization applications require high resolution traffic, vehicle telemetry, weather and road network data. As image processing methods have continued to progress, video streams offer potential insights to user behavior and vehicle occupancy.

There are numerous challenges in storing and processing data for AI-driven urban mobility systems. First, these sources present data in domain-specific formats and at irregular intervals that can vary by provider and source, making it challenging to join data streams to be used by downstream AI models \cite{wang2000spatiotemporal}. Second, the spatiotemporal nature of these data sources presents challenges in efficient storage, synthesis and data retrieval \cite{aydin2016modeling}, \cite{wang2019integrated}. A third challenge is efficiently presenting the data to AI researchers and transit experts for data exploration \cite{thudt2013visits}. In addition, there are the typical challenges of working with high-velocity, high-volume streaming data. 

Therefore in this work we present an integrated data management and processing framework for intelligent urban mobility systems that is currently in use by the Chattanooga Area Regional Transportation Agency (CARTA). We discuss the available data sources and our experiences with joining the various data streams for our set of AI driven applications. We also discuss our methods for monitoring the integrity of the data and present a set of publicly available visualization dashboards designed for our transit agency partners. Lastly, we discuss how these tools are currently being used for AI-driven urban mobility applications in Chattanooga. Through our partnership with transit agencies, we are making these tools open-source and providing access to the visualization dashboards and data sets at \cite{smarttransit}.


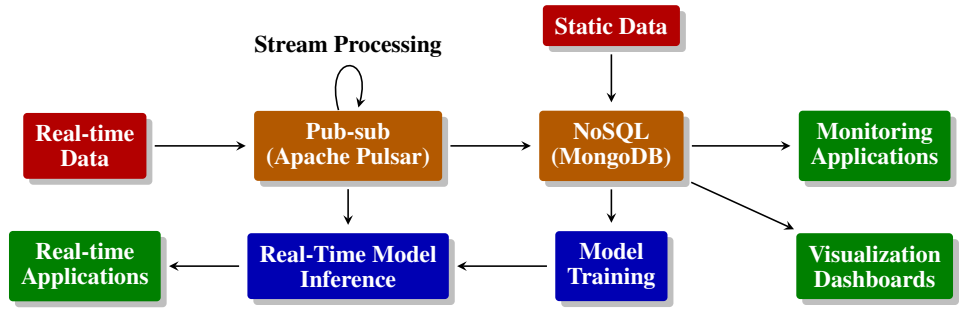
\begin{figure*}[t]
\centering
\begin{tikzpicture}
[
    >=stealth,
    semithick,
    font=\small\bfseries,
    alg/.style = {draw, align=center, font=\linespread{0.95}\selectfont\footnotesize, rounded corners=0.15em, drop shadow=gray, inner sep=0.6em},
    Node/.style={align=center, font=\linespread{0.95}\selectfont\footnotesize\bfseries, rounded corners=0.15em, drop shadow=gray, inner sep=0.6em},
  DataNode/.style={Node, text=white, fill=red!70!black, inner sep=0.16cm, minimum width=1.2cm},
  Arrow/.style={->,black, shorten <= 0.1cm, shorten >= 0.1cm},
  ]
 \tikzset{node distance = 1.6cm} 
    \begin{scope}[node distance=3.5cm]
        \node (n2) [DataNode] {Real-time\\Data};
        \node (n3) [DataNode, fill=orange!70!black,right of=n2]  {Pub-sub \\ (Apache Pulsar)};
        \node (n4) [DataNode, fill=orange!70!black,right of=n3]  {NoSQL \\ (MongoDB)};
        \node (n8) [DataNode, fill=green!50!black,right of=n4] {Monitoring \\ Applications};
    \end{scope}
    \node (n1) [DataNode,above of=n4] {Static Data};
    \node (n7) [DataNode, fill=green!50!black,below of= n2]  {Real-time \\ Applications};
    \node (n9) [DataNode, fill=green!50!black,below of= n8]  {Visualization \\ Dashboards};
    
    \begin{scope}[node distance=3.5cm]
    
    \node (n6) [DataNode, fill=blue!70!black,right of=n7]  {Real-Time Model\\Inference};
    \node (n5) [DataNode, fill=blue!70!black,right of=n6] {Model\\Training};
    \end{scope}

    \draw[Arrow] (n2) --  node[black,midway,right,sloped,font=\scriptsize]{} (n3);
    \draw[Arrow] (n3) --  node[black,midway,right,sloped,font=\scriptsize]{} (n4);
    \draw[Arrow] (n4) --  node[black,midway,below,sloped,font=\scriptsize]{} (n5);
    \draw[Arrow] (n5) --  node[black,midway,left,sloped,font=\scriptsize]{} (n6);
    \draw[Arrow] (n6) --  node[black,midway,left,sloped,font=\scriptsize]{} (n7);
    \draw[Arrow] (n1) --  node[black,midway,below,sloped,font=\scriptsize]{} (n4);
    \draw[Arrow] (n3) --  node[black,midway,below,sloped,font=\scriptsize]{} (n6);
    \draw[Arrow] (n4) --  node[black,midway,below,sloped,font=\scriptsize]{} (n8);
    \draw[Arrow] (n4) --  node[black,midway,below,sloped,font=\scriptsize]{} (n9);
    \path (n3) edge [loop above] node {Stream Processing} (n3);
    
  \end{tikzpicture}
  \caption{Data architecture overview - real time data is streamed to an Apache Pulsar cluster consisting of 5 broker/bookie nodes and 5 zookeeper nodes running on-site in VMWare. A MongoDB cluster running in Google Cloud reads from the Pulsar cluster, continuously updating its data view and adding spatial indexing for monitoring and dashboard applications.}
\label{fig:data_architecture_overview}
    
\end{figure*}

\section{Data Sources} \label{section: data sources}

In this section we outline the available real-time and static data sources. A summary of the available data sources is provided in table \ref{table: data sources}.

\begin{table*}[]
\centering
\caption{Data sources.}
\resizebox{0.9\textwidth}{!}{%
\begin{tabular}{|l|l|l|l|l|l|}
\hline
\textbf{Data} & \textbf{Source} & \textbf{Frequency} & \textbf{Scope} & \textbf{Features} & \textbf{Schema/Format} \\ \hline
\textbf{Diesel vehicles} & ViriCiti and Clever Devices & 1 Hz & 50 vehicles & \begin{tabular}[c]{@{}l@{}}GPS, fuel-level, fuel rate, \\ odometer, trip ID, driver ID\end{tabular} & Viriciti SDK and Clever API \\ \hline
\textbf{Electric vehicles} & ViriCiti and Clever Devices & 1 Hz & 3 vehicles & \begin{tabular}[c]{@{}l@{}}GPS, charging status, battery current, \\ voltage, state of charge, odometer\end{tabular} & Viriciti SDK and Clever API \\ \hline
\textbf{Hybrid vehicles} & Viriciti and Clever Devices & 1 Hz & 7 vehicles & \begin{tabular}[c]{@{}l@{}}GPS, fuel-level, fuel rate, odometer, \\ trip ID, driver ID\end{tabular} & Viriciti SDK and Clever API \\ \hline
\textbf{Traffic} & HERE and INRIX & 1 Hz & Chattanooga and Nashville region & \begin{tabular}[c]{@{}l@{}}TMC ID, free-flow speed, \\ current speed, jam factor, confidence\end{tabular} & \begin{tabular}[c]{@{}l@{}}Traffic Message Channel \\ (TMC)\end{tabular} \\ \hline
\textbf{Road network} & OpenStreetMap & Static & Chattanooga and Nashville region & Road network map, network graph & \begin{tabular}[c]{@{}l@{}}OpenStreetMap \\ (OSM)\end{tabular} \\ \hline
\textbf{Weather} & DarkSky & 0.1 Hz & Chattanooga and Nashville region & \begin{tabular}[c]{@{}l@{}}Temperature, wind speed, \\ precipitation, humidity, visibility\end{tabular} & Darksky API \\ \hline
\textbf{Elevation} & \begin{tabular}[c]{@{}l@{}}Tennessee\\ GIC\end{tabular} & Static & Chattanooga region & Location, elevation & GIS - Digital Elevation Models \\ \hline
\textbf{\begin{tabular}[c]{@{}l@{}}Fixed-line transit \\ schedules\end{tabular}} & CARTA, WeGO & Static & Chattanooga and Nashville region & \begin{tabular}[c]{@{}l@{}}Scheduled trips and trip times, \\ routes, stops\end{tabular} & \begin{tabular}[c]{@{}l@{}}General Transit Feed Specification \\ (GTFS)\end{tabular} \\ \hline
\textbf{Video Feeds} & CARTA & 30 Frames/Second & \begin{tabular}[c]{@{}l@{}}All fixed-line\\ vehicles\end{tabular} & Video frames & Image \\ \hline
\textbf{APC Ridership} & CARTA , Wego & Every Stop & \begin{tabular}[c]{@{}l@{}}All fixed-line \\ vehicles\end{tabular} & \begin{tabular}[c]{@{}l@{}}Passenger boarding count \\ per stop\end{tabular} & Transit authority specific \\ \hline
 
\end{tabular}%
}
\label{table: data sources}
\end{table*}

\subsection{Real-time Data}

CARTA's vehicle fleet for the fixed-line bus transit system includes 50 diesel, 3 electric and 7 hybrid vehicles. Each vehicle has a telematics kit produced by ViriCiti LLC that provides real time telemetry data at a minimum of 1Hz resolution of all available vehicle operating parameters. In total, we have already obtained around 32.3 million data points for electric buses and 29.8 million data points for diesel buses. The nature of the telemetry data is dependent on the type of vehicle. For instance diesel and hybrid vehicles include fuel level and fuel rate where the electric vehicles monitor state-of-charge. All vehicles include GPS and odometer data. Each data reading from ViriCiti includes the label (i.e. GPS), a timestamp and a unique vehicle ID. We collect this telemetry data in real-time from the ViriCiti API \cite{viricitisdk}.

Additionally, each vehicle is equipped with a kit from Clever Devices \cite{cleverdevices}. This data includes GPS, unique vehicle ID (which corresponds with the vehicle ID from ViriCiti) and additionally includes a unique driver ID and the unique trip ID which that vehicle is serving. The unique vehicle ID maps directly to the GTFS schedule produced by CARTA.

We also collect weather data from multiple weather stations in Chattanooga at 5-minute intervals using the DarkSky API. This data includes real-time temperature, humidity, air pressure, wind speed, wind direction, and precipitation. In addition, we collect traffic data at 1-minute intervals using the HERE API, which provides speed recordings for segments of major roads, which provides data in the form of timestamped speed recordings from selected roads. Every road segment is identified by a unique Traffic Message Channel identifier (TMC ID). Each TMC ID is also associated with a list of latitude and longitude coordinates, which describe the geometry of the road segment. Lastly, vehicles are currently being fitted with video equipment that generates real-time video streams to help monitor capacity requirements. 

\subsection{Static Data Sources}

Road network map data was collected from OpenStreetMaps \cite{haklay2008openstreetmap}, which provides road infrastructure modeled as a graph. In addition, we collect static GIS elevation data from the Tennessee Geographic Information Council \cite{tnelevationdata}. From this source, we download high-resolution digital elevation models (DEMs), derived from LIDAR elevation imaging, with a vertical accuracy of approximately 10 cm. We incorporated the elevation data in the OSM network by adding the elevation from the GIS data to each node in the OSM network. Lastly, the vehicle scheduling information is provided by the CARTA in GTFS format.

\section{Data Management} \label{section: data management}

Given the volume and the rate of the data being collected, we had to design a custom architecture for the project. The purpose of this architecture is to store the data streams in a way that provides easy access for offline model training and updates as well as real-time access for system monitoring prediction. This architecture consists of a publish-subscribe cluster implemented with Apache Pulsar, which stores topic-labeled sensor streams, and a MongoDB database backend. An overview of the data architecture is provided in figure \ref{fig:data_architecture_overview}. 

This architecture solves two challenges. The first challenge is the persistent storage of the high-velocity, high-volume data streams.  The second challenge is that the data is highly unstructured and irregular and different data streams have to be synchronized and joined efficiently. With this architecture, we stream each data source to a topic-based publish-subscribe (pub-sub) layer that persistently stores each data stream as a separate topic. Further, we used a three-tiered naming convention for topic labeling. The first tier represents the name of the data tenant and all authentication and access is managed at this level. The second tier is the data category, i.e., vehicle telemetry, traffic, weather, etc. The third tier is the topic name, which represents the data source or provider, such as ViriCiti, HERE, or DarkSky. For ViriCiti, the fleet name is appended to the topic name to separate electric, diesel, and hybrid vehicles. The tenant, category, and topic names together form a topic, which downstream applications can use to access the data streams. We persistently store all messages on each topic in an append-only ledger. Therefore, the topic can be used to read data in near real-time or to playback previous data streams to synchronize new downstream applications. All replication is handled at the ledger level, which allows downstream storage and applications to adapt and expand without concern for data resiliency. For this system we used Apache Pulsar \cite{apachepulsar} due to its native support for authentication and access at the tenant level and high throughput. We run Pulsar on-site on a VMWare cluster. 

We include two methods for long term, structured access to the data streams. First, Pulsar includes support for Presto SQL which is a distributed SQL query engine for big data \cite{prestosql}. Presto SQL integrates with the Pulsar data stores to provide an SQL interface on top of the Pulsar topics. This is useful for analytics teams comfortable with SQL, however as it is designed for large scale batch queries and does not support geospatial indexing it is not optimal for user-centric applications such as visualization dashboards. Therefore, we implemented a downstream MongoDB \cite{mongodb} cluster running in Google Cloud. MongoDB was chosen for its native support of geospatial, r-tree indexing which optimizes our system for aggregate geospatial queries for monitoring and visualization applications discussed in the next section.

\section{Data Synthesis and Stream Processing} \label{section: data synthesis and stream processing}

The various downstream applications such as monitoring systems, visualization dashboards and energy and ridership prediction models require data from various streams to be merged. Typical implementations of stream processing architectures require external processing frameworks such as Apache Spark and Storm. For our implementation we decided to join the data streams within Apache Pulsar. This process involves designing functions that read from a set of data stream topics, merge the streams in a series of time windows, and output the joined data on a new Pulsar topic. 

\begin{figure}[t]
\centering
\begin{tikzpicture}
[
    >=stealth,
    semithick,
    font=\small\bfseries,
    alg/.style = {draw, align=center, font=\linespread{0.95}\selectfont\footnotesize, rounded corners=0.15em, drop shadow=gray, inner sep=0.6em},
    Node/.style={align=center, font=\linespread{0.95}\selectfont\footnotesize\bfseries, rounded corners=0.15em, drop shadow=gray, inner sep=0.6em},
  DataNode/.style={Node, text=white, fill=red!70!black, inner sep=0.16cm, minimum width=1.2cm},
  Arrow/.style={->,black, shorten <= 0.1cm, shorten >= 0.1cm},
  ]
 \tikzset{node distance = 1.35cm} 
    \begin{scope}[node distance=3.0cm]
    \node (n12) [DataNode]  {\underline{Input Streams} \\
    - Viriciti \\
    - CleverDevices \\
    - DarkSky \\
    - HERE (traffic) \\
    - Video Feed};
    \node (n22) [DataNode, fill=orange!70!black, right of=n12] {Merge \\ Function};
        \node (n32) [DataNode, fill=green!50!black,right of=n22] {\underline{Output Stream} \\
        - Vehicle ID \\
        - Trip ID \\
        - Route ID \\
        - SOC \\
        - OSM Segment \\
        - Timestamp \\
        - Temperature \\
        - Humidity \\
        - Elevation};
    \end{scope}

    \node (n21) [DataNode, fill=blue!70!black,above of=n22] {\underline{MongoDB Query} \\ Static Data};
    
    \draw[Arrow] (n12) --  node[black,midway,right,sloped,font=\scriptsize]{} (n22);
    \draw[Arrow] (n21) --  node[black,midway,right,sloped,font=\scriptsize]{} (n22);
    \draw[Arrow] (n22) --  node[black,midway,right,sloped,font=\scriptsize]{} (n32);

  \end{tikzpicture}
  \caption{An example stream data join. Real-time telemetry and routing data from CleverDevices and Viriciti is combined with weather from DarkSky, traffic from HERE and the video feed. The output stream includes all fields from these sources, as well as static data from OSM, GTFS and elevation. The output stream is a sliding time window at 5 second intervals.}
\label{fig:merge}
    
\end{figure}
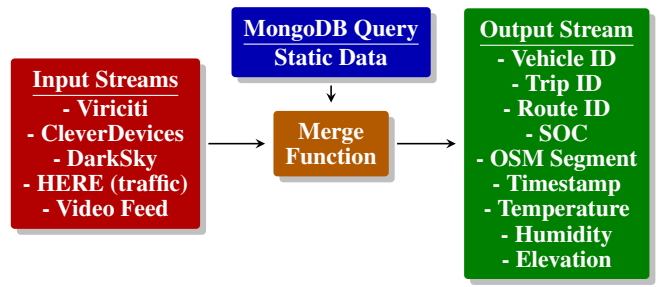

As our framework has expanded, we are running numerous streaming join functions within Pulsar. An example is provided in figure \ref{fig:merge}, which outputs a data stream that is used for our energy prediction models and energy dashboard. The input is the telemetry data from Viriciti, route, trip and driver data from Clever Devices, weather from DarkSky, traffic from HERE and the video feeds. Additionally, our predictive models rely road level information from OSM. As this data is static the latest OSM network is stored in a MongoDB collection which the function queries each evening to keep up-to-date. These data sources are merged at 1 second time windows, which is the resolution required by the predictive models. 

Running our stream processing applications within Pulsar has two benefits. First it provides real-time access to consumers that subscribe to the output topic of these applications. Second, we include a subscriber that continuously adds geospatial indexing the the streams and writes to MongoDB. One disadvantage of this approach is that there is some additional overhead regarding development time working with the Pulsar API's instead of more mature, streaming specific frameworks. However since we do not require the overhead of incorporating a separate stream processing framework we are able to reduce complexity of the overall system and reduce costs that would be associated with running this external system.

\section{Data Monitoring} \label{section: data monitoring}

The incoming data, particularly vehicle telemetry data is provided by cutting edge telemetry kits from our partner companies ViriCiti and Clever Devices. These kits are installed on a variety of fleets bought over the last 25 years, each with different specifications and requirements. The challenges associated with this work require careful monitoring to ensure the quality of the incoming data. Additionally, it is useful to monitor our data architecture itself to identify gaps in coverage or failures in the system. Therefore we implemented a custom monitoring system to notify our data management team and CARTA of when issues arise.

Our monitoring includes automated programs which send nightly emails summarizing the state of the system as well as the incoming data. We use historical data regarding the number of messages on each topic per day of the week to compare with the number of messages on that topic over the previous day. If the number of messages over the previous day was more than two standard deviations less than expected, an email is triggered to notify us there was a discrepancy on one of the data streams. This application runs on all registered topics within Pulsar. 

We found that this monitoring application was insufficient regarding the ViriCiti data since failures with the telemetry kits were highly correlated between vehicle models. For instance, early in this project there were issues regarding the kits for the 10 diesel vehicles bought in the late 1990s. Since these had issues immediately, they never showed up in the historical averages and thus were not identified as an issue. Therefore, a second application was designed specifically for the ViriCiti data. This application queries all vehicles, identified by unique vehicle ID, that serviced a trip that day. We then query the ViriCiti data to ensure there was telemetry data on that vehicle during that time window. 

\section{Visualization Dashboards} \label{section visualization dashboards}

\begin{figure}[t]
\centering
\begin{subfigure}{0.9\columnwidth}
    \includegraphics[width=\textwidth]{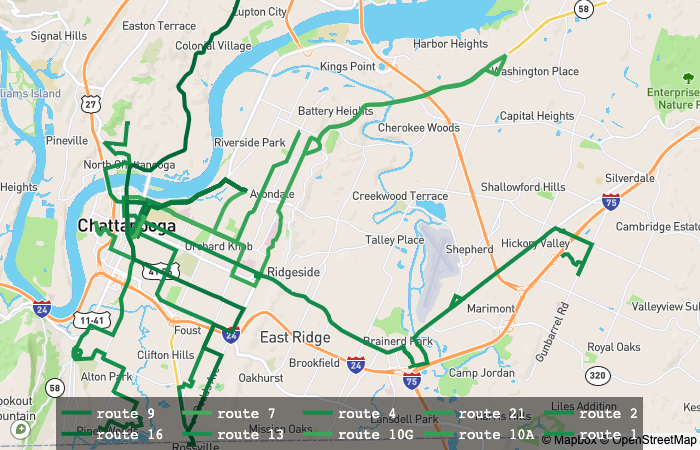}
    \caption{}
    \label{subfig: energy visualization}
\end{subfigure}
\begin{subfigure}{0.9\columnwidth}
    \includegraphics[width=\textwidth]{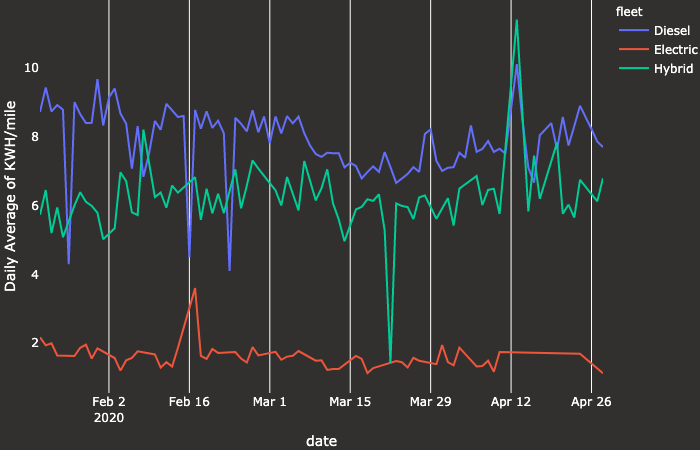}
    \caption{}
    \label{subfig: energy per fleet}
\end{subfigure}
\caption{Energy visualization dashboard: (a) energy consumed per route for electric vehicles from January 1, 2020 to May 1, 2020. (b) Energy consumption per fleet between January 1, 2020 to May 1, 2020. Energy measured in kWh/Mile.}
\label{fig: dashboard}
\vspace{-0.2in}
\end{figure}

As our framework developed, we also implemented a set of visualization web dashboards using Python, Plotly and Dash. These dashboards are connected to our MongoDB backend to query data for presentation to the user. An example of our energy visualization dashboard is shown in figure \ref{fig: dashboard}. In this dashboard, the user can query based on time, fleet and route. The data is presented to the user over the map of Chattanooga as shown in figure \ref{subfig: energy visualization} and as a series of statistical visualizations, one of which is energy per fleet as shown in figure \ref{subfig: energy per fleet}. This dashboard is used by the data management team and CARTA to monitor the performance of the CARTA fleets over time and is available to the public \cite{energydashboard}. Additionally, we developed a ridership dashboard to visualize occupancy of the vehicles throughout the bus transit network. The presentation of the occupancy dashboard is similar to the energy dashboard, and is available at \cite{occupancydashboard}.


\section{AI Applications} \label{section: ai applications}

In addition to the visualization dashboard applications, we are currently running a set of AI applications that rely upon the data management framework and data sources discussed in this paper. The first of which is an energy prediction model presented in \cite{ayman2020data}. These prediction models use the output features as shown in figure \ref{fig:merge} to train regression and neural network models to predict the energy that will be consumed on a route by the diesel, hybrid and electric vehicles. We are currently working on incorporating these models in the energy dashboard to help CARTA with vehicle scheduling and operational guidance. 

Additionally, we are working training statistical models to predict vehicle occupancy to better schedule vehicles in the context of social distancing regulations from COVID-19. In this way we can help to ensure these safety requirements are met and help CARTA better schedule vehicles on popular routes. These models will be incorporated in the ridership dashboards so CARTA operators have real-time access to these models.

\section{Conclusion and Future Work}

In this work we presented our integrated data management and processing framework for intelligent urban mobility systems, which is currently in use by our partner transit agencies. We also covered our associated monitoring systems, visualization dashboards and briefly discussed the current AI applications using these tools which we are making open-source and providing access to the visualization dashboards and data sets at \cite{smarttransit}.

In future work we are interested in investigating decentralized edge-cloud hybrid architectures for AI-driven urban mobility systems. We have done some work on AI-driven, decentralized routing applications using federated learning \cite{wilbur2020time} and fog-cloud middleware for smart mobility systems \cite{talusan2019smart}, \cite{talusan2020decentralized}. We are interested in studying problems related to data storage and retrieval for these systems in future work.

\section{Acknowledgement}

This material is based upon work supported by the Department of Energy, Office of Energy Efficiency and Renewable Energy (EERE), under Award Number DE-EE0008467 and National Science Foundation  through award numbers 1818901, 1952011, 2029950 and 2029952. The authors will also like to acknowledge the computation resources provided by the Research Computing Data Core at the University of Houston and through cloud research credits provided by Google. 
The material presented here presents the views of the authors and do  not necessarily state or reflect those of the United States Government or any agency thereof. Neither the United States Government nor any agency thereof, nor any of their employees, makes any warranty, express or implied, or assumes any legal liability or responsibility for the accuracy, completeness, or usefulness of any information, apparatus, product, or process disclosed, or represents that its use would not infringe privately owned rights. Reference herein to any specific commercial product, process, or service by trade name, trademark, manufacturer, or otherwise does not necessarily constitute or imply its endorsement, recommendation, or favoring by the United States Government or any agency thereof.




\balance
\bibliographystyle{aaai}
\bibliography{main}

\end{document}